\documentclass[twocolumn,showpacs,prl,amsmath,amssymb]{revtex4}

\usepackage{graphicx}
\usepackage{dcolumn}
\usepackage{bm}
\usepackage{color}
\begin{document}
\title{Thermal equilibrium of non-neutral plasma in dipole magnetic field}
\author{N. Sato, N. Kasaoka and Z. Yoshida}
\affiliation{Graduate School of Frontier Sciences, The University of Tokyo,
Kashiwa, Chiba 277-8561, Japan}
\date{\today}

\begin{abstract}
Self-organization of a long-lived structure is one of the remarkable characteristics of macroscopic systems 
governed by long-range interactions.
In a homogeneous magnetic field, a non-neutral plasma creates a ``thermal equilibrium'' 
which is a Boltzmann distribution on a rigidly rotating frame.
Here, we study how a non-neutral plasma self-organizes in inhomogeneous magnetic field; 
as a typical system we consider a dipole magnetic field.
In this generalized setting, the plasma exhibits its fundamental mechanism that determines the relaxed state.
The scale hierarchy of adiabatic invariants is the determinant;
the Boltzmann distribution under the topological constraint by the robust adiabatic invariants
(hence, the homogeneous distribution with respect to the fragile invariant)
is the relevant relaxed state,
which turns out to be a rigidly rotating clump of particles
(just same as in a homogeneous magnetic field),
while the density is no longer homogeneous.
\end{abstract}


\maketitle


\section{Introduction}
Self-organization of a long-lived structure (inhomogeneity of physical quantities) is often
observed in macroscopic systems governed by long-range interactions
such as gravity (creating astronomical systems like galaxies\,\cite{Lynden-Bell}), 
electromagnetic force (creating plasma systems like magnetospheres\,\cite{Schulz,Boxer,yoshida2013} 
or particle traps\,\cite{Penning}), 
or magnetic interaction (Hamiltonian mean-field systems modeling magnetism\,\cite{Antoniazzi,Pakter}).
The common physics is described by the Vlasov equation coupled with a relevant field equation.
The long-lived structure is a particular stationary solution that is ``robust'' against microscopic perturbations.
Here, we put the problem into the perspective of non-canonical Hamiltonian mechanics\,\cite{Morrison},
and show that the self-organization occurs on a leaf of topologically constrained phase space;
the topological constraint originates from the adiabatic invariants, which defines a macroscopic hierarchy\,\cite{YoshidaMahajan2014}.
As a specific system, we consider a non-neutral (single species) plasma in a dipole magnetic field.
Let us begin by explaining how this problem is interesting from both basic and applied physics viewpoints.

When a non-neutral plasma is put in a homogeneous longitudinal magnetic field, 
it is spontaneously confined, relaxing into a ``thermal equilibrium'' on a rigidly rotating frame\,\cite{Penning,Mal1,Mal2}.
Canceling the self electric field by the Lorentz-transformed electric filed,
the Boltzmann distribution yields a homogeneous density profile inside the confinement region.
The constant density profile and the constant angular momentum profile are the simultaneous characteristics of the 
\emph{relaxed state}; there remains no free energy to excite macroscopic perturbations
(as far as the system conserves the total angular momentum and is
isolated from other energy sources like other species of particles or external electromagnetic fields). 
However, the consistency of the density distribution 
(dictated by the statistical mechanics of particles)
and the electric potential (dictated by the field equation)
relies heavily on the specialty of the homogeneous longitudinal magnetic field.
Here, we investigate whether such a relaxed state exists in an inhomogeneous magnetic field.
Experimentally it does exist in a dipole magnetic field\,\cite{yoshida2010,saitoh2010};
the charged particles self-organize a rigidly rotating clump.
However, the density is no longer homogeneous.
The aim of this study is to
reveal the underlying principle that governs generalized relaxed states.
Confinement of charged particles (especially antimatter particles) in a toroidal magnetic bottle
has many advantages, for example, making possible to confine high-energy particles produced by isotopes or
accelerators, or to confine different spices of positive and negative charges simultaneously
\cite{Yoshida1999,Pedersen1,Pedersen2}.

\begin{figure}[tb]
\hspace*{-0.9cm}
\includegraphics[scale=0.2]{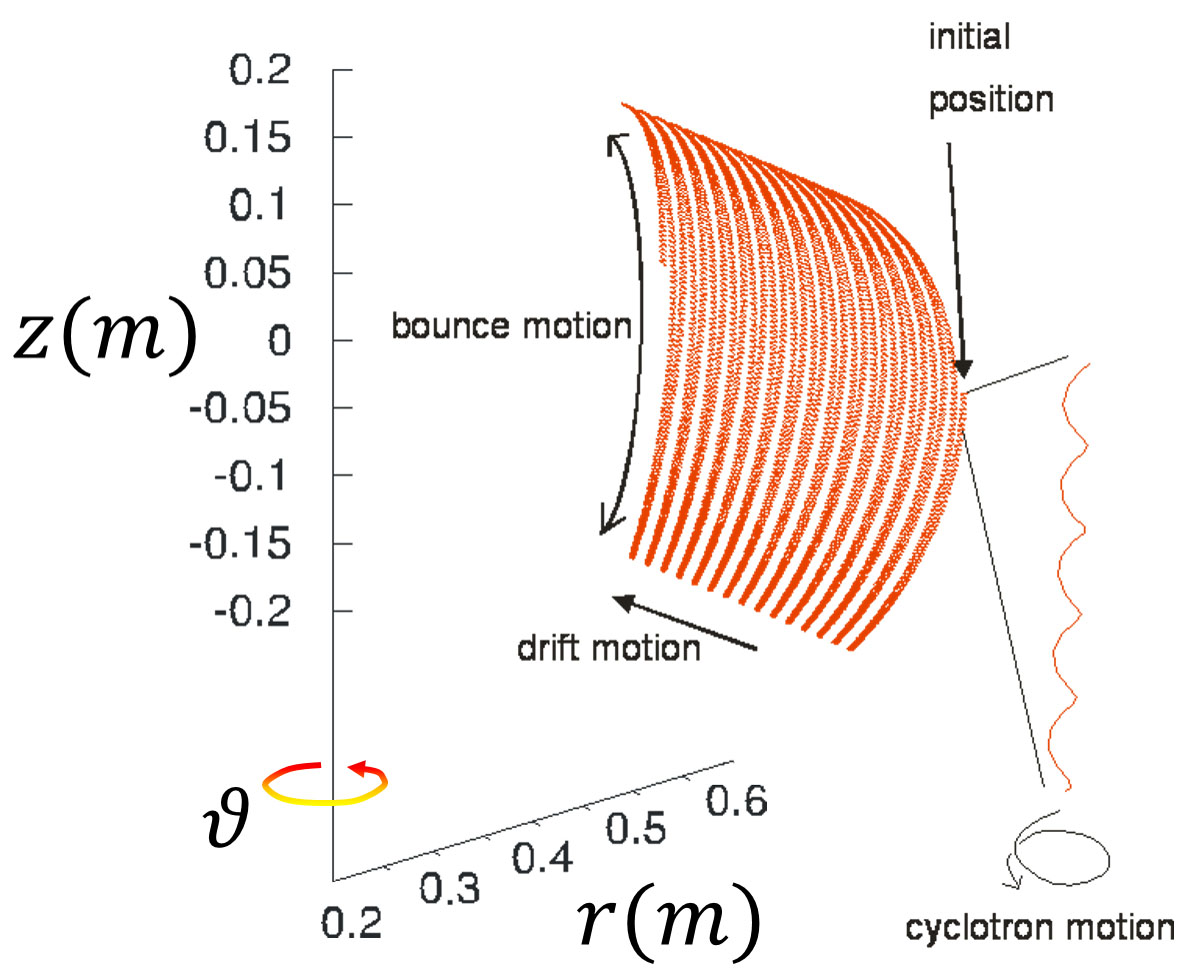}
\caption{\footnotesize A typical orbit of a magnetized particle in a dipole magnetic field,
which has a hierarchy of three different frequencies.
Here we assume the parameters of the RT-1 device \cite{yoshida2010}; the magnetic field at the confinement region is about 0.05T.
For a particle having an isotropic energy of 50eV,
$\omega_c \sim$ 1GHz, $\omega_b\sim$ 1MHz, 
and $\omega_d\sim$ 1kHz.}
\label{orbit}
\end{figure}

Here we invoke the theory of phase-space foliation (or, topological constraint), 
and define a relaxed state as a thermal equilibrium on a leaf of phase space\,\cite{YoshidaMahajan2014}.
In the present argument, the adiabatic invariants of magnetized particles embody such foliated phase space.
In an axisymmetric magnetic field,
magnetized particles have three different adiabatic invariants, 
i.e., the magnetic moment $\mu$, the action $J_\parallel$ of bounce motion, 
and the action (canonical angular momentum) $P_\theta$ of the toroidal drift\,\cite{adiabatic};
see Fig.\,\ref{orbit}.
We may approximate $P_\theta/q$ ($q$ is the charge) by the magnetic flux function $\psi$ such that $\bm{B}=\nabla\psi\times\nabla\theta$ ($\theta$ is the toroidal angle).
When the magnetic field is sufficiently strong, the corresponding frequencies define a hierarchy:
$\omega_c$ (cyclotron frequency) $\gg$ $\omega_b$ (bounce frequency) $\gg$ $\omega_d$ (drift frequency).
Hence, $\psi$ is the most fragile constant
---the homogenization with respect to $\psi$ yields the relaxed state on the first (macroscopic) hierarchy of the adiabatic invariants.
Needless to say, the ultimate relaxed state is achieved at the maximum homogeneity (maximum entropy)
after destroying all adiabatic invariants, and it is the thermal death.


We define the ``relaxed state'' by a distribution function $f$ that has no $\psi$ dependence,
i.e., 
\begin{equation}
\frac{\partial f }{\partial \psi}=0.
\label{relaxed-state_definition}
\end{equation}
For the equilibrium to be non-trivial, we demand 
that the total canonical angular momentum $\int \psi f d^6z$ to be a non-zero constant
($d^6z$ denotes the volume element of the phase space).

\section{Kinetic model of macroscopic relaxed sate}

In order to formulate the model with taking into account the hierarchy of adiabatic invariants,
we write the Hamiltonian of a particle as
\begin{equation}
H_{gc} = \omega_c \mu + \omega_b J_\parallel + q \phi .
\label{GCHamiltonian}
\end{equation}
We have omitted the kinetic energy $(P_\theta-q\psi)^2/(2mr^2)$ 
of the toroidal drift velocity by approximating $P_\theta=q\psi$\,\cite{drift_kinetic}.
By the symmetry, the toroidal angle $\theta$ is not included in $H_{gc}$.
The gyro angle ${\vartheta_c}$ (which is conjugate to the magnetic moment $\mu$) is coarse-grained
by replacing $\dot{\vartheta_c}$ with $\omega_c$, and is completely eliminated from $H_{gc}$
(i.e., $H_{gc}$ dictates the motion of the guiding center of the gyrating particle).  
However, the bounce angle ($\vartheta_b$) is not ignored,
because the frequencies $\omega_c$ and $\omega_b$, as well as the electric potential $\phi$ 
are functions of the spacial coordinates including $\vartheta_b$.
Here we choose $\psi$ and $\zeta$ (the parallel coordinate along each magnetic surface, the
level-set of $\psi$) as the spatial coordinates 
(then, $\vartheta_b=\pi\zeta/\ell_\parallel$; $\ell_\parallel$ is the bounce orbit length).

The action $J_\parallel$ is conjugate to $\vartheta_b$:
\begin{equation}
\dot{J}_\parallel=\frac{\partial H_{gc}}{\partial \vartheta_b} 
=\frac{\ell_\parallel}{\pi} \frac{\partial H_{gc}}{\partial\zeta}.
\label{bounce_action}
\end{equation}
For the periodic bounce motion, 
$\oint (\partial H_{gc}/\partial \vartheta_b)d\vartheta_b =\oint d H_{gc}=0$.
Integrating (\ref{bounce_action}) over the cycle of bounce motion yields
the bounce-average $\langle J_\parallel\rangle=$ constant.
When we calculate macroscopic quantities (like the total energy or the total action),
we evaluate $J_\parallel$ as the adiabatic invariant $\langle J_\parallel\rangle$.

The drift frequency (including all grad-B, curvature, and E$\times$B drifts) is given by
bounce-averaging the toroidal angular velocity
\begin{equation}
\omega_d=\dot{\theta} = \frac{\partial H_{gc}}{\partial\psi} 
=\mu \frac {\partial\omega_c }{\partial \psi}+ J_\parallel\frac{\partial\omega_b }{\partial \psi} 
+ q \frac{\partial \phi}{\partial\psi}.
\label{drift_frequency}
\end{equation}
In a homogeneous magnetic field, both $\omega_c$ and $\omega_b$ are constant, and then
(\ref{drift_frequency}) evaluates the  E$\times$B drift frequency.

In terms of the constants of motion $H_{gc}$, $\mu$, $J_\parallel$, and $\psi$,
a general equilibrium solution of the drift kinetic equation (such that $\{H_{gc},f\}=0$) 
is written as $f(H_{gc},\mu,J_\parallel,\psi)$.
The relaxed state is the special solution that maximizes the entropy
$S=-\int f \log f \,d^6z$ under the constraints on
\begin{enumerate}
\item the total particle number $N= \int f\,d^6z$,
\item the total energy $E=\int H_{gc} f\,d^6z$,
\item the total magnetic moment $C_\mu=\int \mu f\,d^6z$,
\item the total bounce action $C_{J_\parallel}=\int J_\parallel f\,d^6z$,
\item the total angular momentum $C_\psi = \int \psi f\,d^6z$.
\end{enumerate}
The variational principle yields
\begin{equation}
f_{T}(H_{gc},\mu,J_\parallel,\psi) 
= Z^{-1} e^{-\beta (H_{gc} - \gamma_1\mu - \gamma_2 J_\parallel - \gamma_3 \psi)},
\label{solution-1}
\end{equation}
where $Z$ (normalization factor), $\beta$ (inverse temperature), 
$\gamma_1$, $\gamma_2$, and $\gamma_3$ are constants
related to the Lagrange multipliers on
$N$, $E$, $C_\mu$, $C_{J_\parallel}$, and $C_\psi$.

While we derived (\ref{solution-1}) for a given set of constants,
we may, alternatively, regard $f_{T}$ as a Boltzmann distribution on a
\emph{grand-canonical ensemble} parametrized by the aforementioned macroscopic quantities,
and then, we interpret $\gamma_1, \gamma_2$ and $\gamma_3$
as the chemical potentials pertinent to the changes in the action variables $\mu$, $J_\parallel$, and $\psi$,
respectively
(remember the parallel relations between the ``energy level'' and the frequency, 
as well as between the ``particle number'' and the action variable, in analogy with the Landau levels in quantum theory).

Finally, the determining equation (\ref{relaxed-state_definition}) must be satisfied,
which is equivalent to
\begin{equation}
\frac{\partial H_{gc}}{\partial \psi} = \gamma_3 ~(= \textrm{constant}) .
\label{ASEquilibrium}
\end{equation}
Remembering (\ref{drift_frequency}), we find that (\ref{ASEquilibrium}) implies rigid rotation. 

The relaxed-state plasma occupies a finite domain that is surrounded by a magnetic surface, 
i.e., we can connect the distribution function $f_{T}$ to the vacuum $f=0$
at some level-set $\psi=\psi^*$.
Invoking Heaviside's step function $Y(\psi^*-\psi)$ 
(which is zero inside the plasma region),
we may write the extended distribution function as
\begin{equation}
\tilde{f}_{T}(H_{gc},\mu,J_\parallel,\psi)
=\lim_{\alpha\rightarrow\infty}e^{-\alpha Y(\psi^*-\psi)}\cdot f_{T},
\label{solution-2}
\end{equation}
which is in equilibrium ($\{H_{gc},\tilde{f}_{T}\}=0$)
and relaxed: $\partial\tilde{f}_{T}/\partial\psi=0$
except at the boundary (the boundary is not ``relaxed'' as it reflects the
constraint given by $C_\psi$; later, we will show how the boundary is determined).

 Neglecting the current generated by the plasma, $\psi$ is a given function.
 Then, (\ref{ASEquilibrium}) may be viewed as a determining equation for the electric potential $\phi$
 (in addition to the term $q\phi$, $J_\parallel$ depends on $\phi$ in a complex way;
 however, if the electric field is much stronger than the thermal energy, we may
 approximate $\bm{B}\cdot\nabla\phi\approx 0$, and then, $J_\parallel$ is independent to $\phi$, and 
 we may put $\phi=\phi(\psi)$).

The electric potential $\phi$ included in $H_{gc}$ must be consistent to the field equation
\begin{subequations}
\begin{align}
-\nabla^2\phi &= 4\pi q \rho,
\label{Poisson_eq}
\\
\rho &= \int f_{T} (\bm{x},\bm{v})\,d^3v
= \int f_{T}\, \frac{2\pi\omega_c d\mu}{m} \frac{dJ_\parallel}{m\ell_\parallel} .
\label{density}
\end{align}
\end{subequations}
The existence of a self-consistent field $\phi$ satisfying both (\ref{ASEquilibrium}) and (\ref{Poisson_eq}) is not at all obvious. In what follows, we will construct non-trivial solutions;
one is  the well-known ``thermal equilibrium'' in a straight homogeneous magnetic field, and the other is a new solution (numerical) in a dipole magnetic field.

\section{Thermal equilibrium in a homogeneous magnetic field}
First, we put the classical solution into the new perspective formulated here.
In a homogeneous longitudinal magnetic field
$\bm{B}=B_0 \nabla z $ ($B_0$ is a constant), $\omega_c $ and $\omega_b$ are constants.
We may put $\psi=B_0r^2/2$ and $\phi=\phi(\psi)$ (assuming that the relaxed state is
homogeneous with respect to $\zeta=z$).
The relaxed-state condition (\ref{relaxed-state_definition}) reads $q\partial\phi/\partial\psi=\gamma_3$,
which yields $\phi=(\gamma_3/q)\psi$, and
\begin{equation}
f_{T} = Z^{-1} e^{-\beta[ (\omega_c-\gamma_1)\mu + (\omega_b-\gamma_2)J_\parallel ]}.
\label{Penning-2}
\end{equation}
Since this distribution function has no spatial dependence, $\rho=\rho_0$ (constant).
On the field equation (\ref{Poisson_eq}),
the potential $\phi = (\gamma_3/q) \psi = (\gamma_3/q) B_0r^2/2$ is
consistent to the flat density $\rho_0$, 
if  $\gamma_3= -2\pi q^2 \rho_0/B_0=-\omega_p^2/(2c\omega_c)$.
The parameter $\gamma_3$ controls the density $\rho_0$
(while $\gamma_1$ and $\gamma_2$ change the velocity anisotropy).
Given the particle number $N$ (per unit length in $z$), $\rho_0=N/(\pi R^2)$, where $R$ is the
radius of the plasma column.
The total angular momentum evaluates 
$C_\psi= B_0 N R^2/4$.
Hence, for given $N$ and $C_\psi$, we obtain $R=2\sqrt{C_\psi/B_0 N}$.

\section{Relaxed state in a dipole magnetic field}

In a dipole magnetic field, both $\omega_c$ and $\omega_b$ vary as functions of the spacial
coordinates $\psi$ and $\zeta$.
Because of the inhomogeneous Jacobian weight in the integral (\ref{density}), the parameters $\gamma_1$ and $\gamma_2$, included in $f_{T}$, cause
a change in the profile of $\rho$\,\cite{hasegawa2005,yoshida2013},
which is in marked contrast to the foregoing case of homogeneous magnetic field.

Assuming some $\phi$ in $f_{T}$,
we calculate $\rho$ by (\ref{density}), and then solve (\ref{Poisson_eq}) for a new $\phi$.
Iterating this process, we obtain a self-consistent $\phi$ and a kinetic equilibrium $f_{T}$.
Next we have to vary the parameters $\gamma_1$, $\gamma_2$, and $\gamma_3$ to find 
the relaxed state.

Since the left-hand side of (\ref{ASEquilibrium}) contains $\mu$ and $J_\parallel$,
we cannot satisfy (\ref{ASEquilibrium}) for each particle having different $\mu$ and $J_\parallel$
(excepting the case when $\partial\omega_c/\partial\psi=\partial\omega_b/\partial\psi=0$,
as it is in a homogeneous magnetic field).
Instead, we demand that the macroscopic drift velocity
\begin{equation}
\overline{v}_{\theta}= \rho^{-1} \int v_{\theta}f_{T} \frac{2\pi\omega_{c}}{m}d\mu dv_{\parallel}
\end{equation}
is rigid rotation ($v_\theta$ is directly evaluated by the orbit calculations).
The chemical potentials $\gamma_1,\gamma_2$, and $\gamma_3$ are the control parameters
to be optimized to yield rigid rotation.

\begin{figure}[tb] 
\hspace*{-0.7cm}
\includegraphics[scale=0.4]{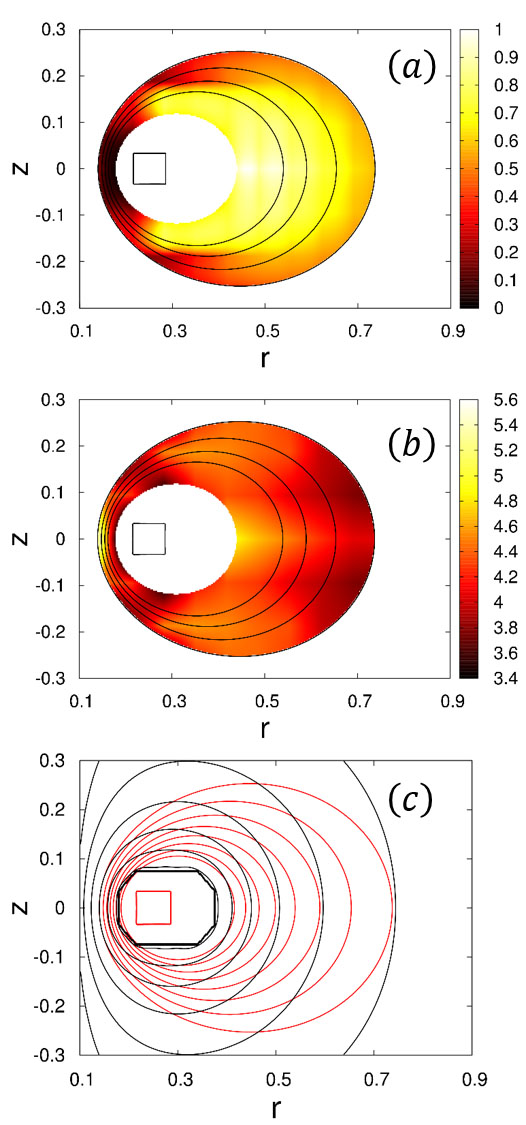}
\caption
{\label{solution} A relaxed state in a dipole magnetic field:
(a) self-consistent density profile $\rho$, (b) the distribution of the angular velocity $\omega_{d}$, and (c) the contours of the electric potential $\phi$ (black lines) and the magnetic flux function $\psi$ (the magnetic field lines; red lines).
}
\end{figure}

Figure\,\ref{solution} shows an example of solution.
Here we assume parameters that simulate the RT-1 experiment~\cite{yoshida2010}.
The temperature $\beta^{-1}=50$eV is chosen to be the typical energy of the injected electrons.
Other parameters are
$(\gamma_{1},\gamma_{2},\gamma_{3})=(-8.0\times10^{-2}\,T,-7.2\times10^{6}\,s^{-1},\,1.4\times10^{4}\,Wb^{-1}\,eV)$,
by which $\beta\gamma_{1}\mu$, $\beta\gamma_{2}J_{\parallel}$, $\beta\gamma_{3}\psi$, and $\beta q\phi$ are all of order $\beta H\sim 1$ (implying that all terms in $f_{T}$ play essential roles in characterizing the relaxed state).

\section{Numerical analysis}
Here we study how general equilibrium solutions (consistent $\phi$ and $f_{T}$
satisfying $\{H_{gc},f_{T}\}=0$ and the Poisson equation (\ref{Poisson_eq}) simultaneously)
vary as the parameters are changed,
and how these parameters can be optimized to send the equilibrium to the relaxed state.

For the convenience, we introduce indexes to evaluate the ``relaxation'':
\begin{subequations}
\begin{align}
&\bar{\omega}_{d}=\frac{1}{N}\int{\omega_{d}\rho d^{3}x}, \\
&\sigma_{\omega_{d}}=\sqrt{\frac{1}{N}\int{\left(\omega_{d}-\bar{\omega}_{d} \right)^{2}\rho d^{3}x}}, \\
&\chi=\frac{\sigma_{\omega_{d}}}{\bar{\omega}_{d}},
\end{align}
\end{subequations}
i.e.,  $\bar{\omega}_{d}$ the spatially averaged toroidal drift
frequency, $\sigma_{\omega_{d}}$ the associated standard deviation, and $\chi$ a measure of rigidity of the rotation.

\begin{figure}[tb]
\hspace*{-0.9cm}
\includegraphics[scale=0.15]{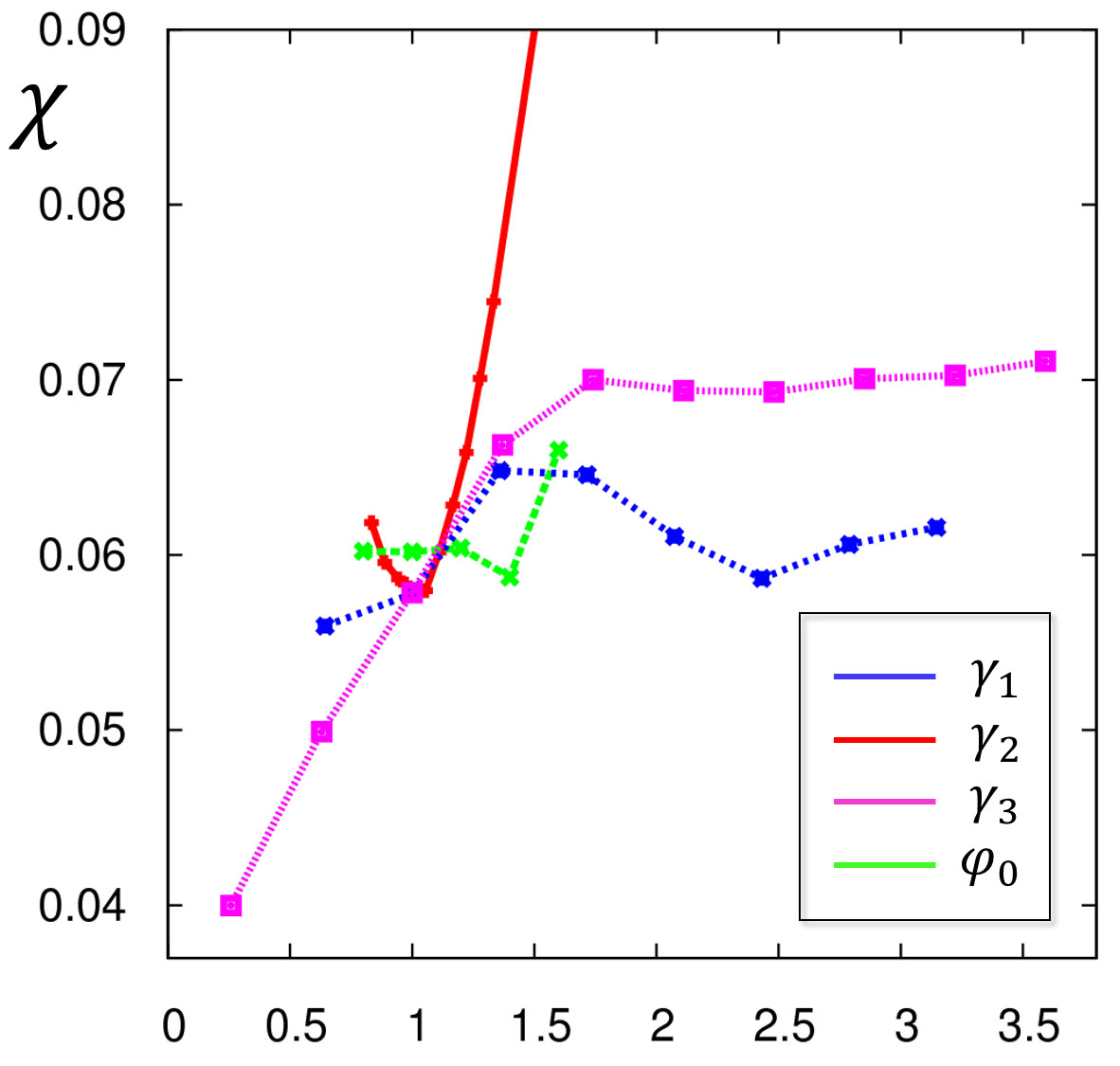}
\caption{ The rigidity parameter $\chi$ as functions of $\gamma_1$, $\gamma_2$, $\gamma_3$ and $\phi_{0}$.
Each parameter is normalized by the corresponding central value
$(\gamma_{1},\gamma_{2},\gamma_{3},\phi_0)=(-8.0\times10^{-2}\,T,-7.2\times10^{6}\,s^{-1},\,1.4\times10^{4}\,Wb^{-1}\,eV,-250\,V)$,
and changed independently.
}
\label{dependences}
\end{figure}

In Fig. \ref{dependences}, the behavior of $\chi$ 
as a function of the control parameters $\gamma_{1}$, $\gamma_{2}$, $\gamma_{3}$, and $\phi_{0}$ (the electric potential at the coil surface) is shown. A detailed explanation for each parameter is given in the following subsections.





\subsection{Chemical potential $\gamma_{1}$ of the magnetic moment $\mu$}

\begin{figure}[h]
 \hspace*{-0.5cm}
 \includegraphics[scale=0.185]{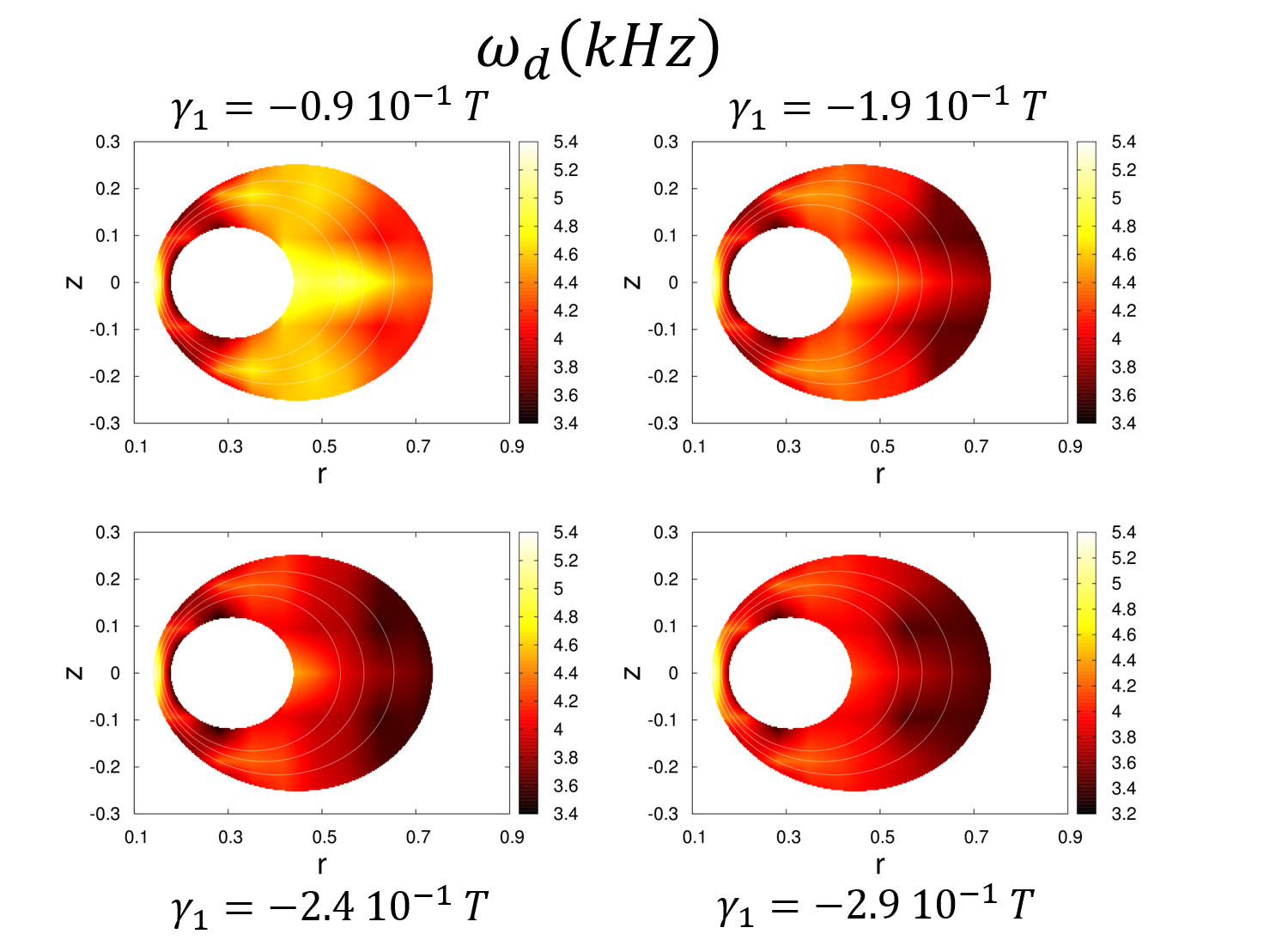}
 \caption{Plots of the angular frequency $\omega_{d}(kHz)$ of rotation for four different values of $\gamma_{1}(T)$.}
 \label{omegadg1}
 \end{figure}

\begin{figure}[h]
 \hspace*{-0.5cm}
 \includegraphics[scale=0.185]{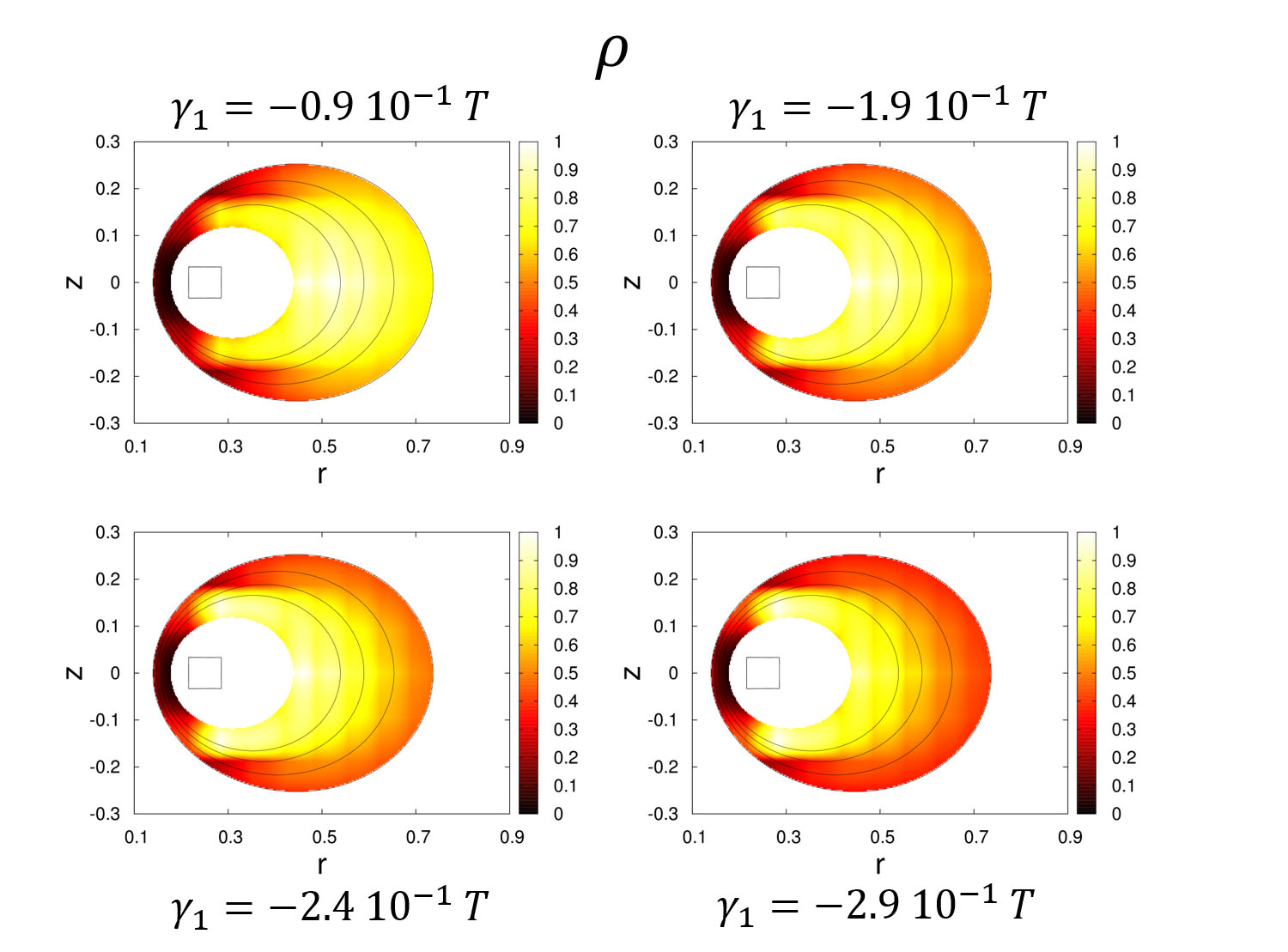}
 \caption{Plots of the density $\rho$ (normalized) for four different values of $\gamma_{1}(T)$.
 }
 \label{rhog1}
 \end{figure}

The dependence of the equilibrium state on the parameter $\gamma_1$
(the chemical potential of the magnetic moment $\mu$) 
is shown in Figs. \ref{omegadg1} (plots of $\omega_d$) and \ref{rhog1} (plots of $\rho$).
Here, other parameters are fixed at $\phi_{0}=-250\,V$, 
$\beta^{-1}=50$eV, and $(\gamma_{2},\gamma_{3})=(-0.72\times10^{7}s^{-1},1.25\times10^{4}\, Wb^{-1}\,eV)$. 
As $\gamma_1$ is  increased, the density approaches to the distribution $\rho\propto B$\,\cite{yoshida2013}.
However, the rigidity of the rotation is a weak function of $\gamma_{1}$ (see Fig. \ref{dependences}).

\subsection{Chemical potential $\gamma_{2}$ of the bounce action $J_\parallel$}

\begin{figure}[h]
 \hspace*{-0.5cm}
 \includegraphics[scale=0.185]{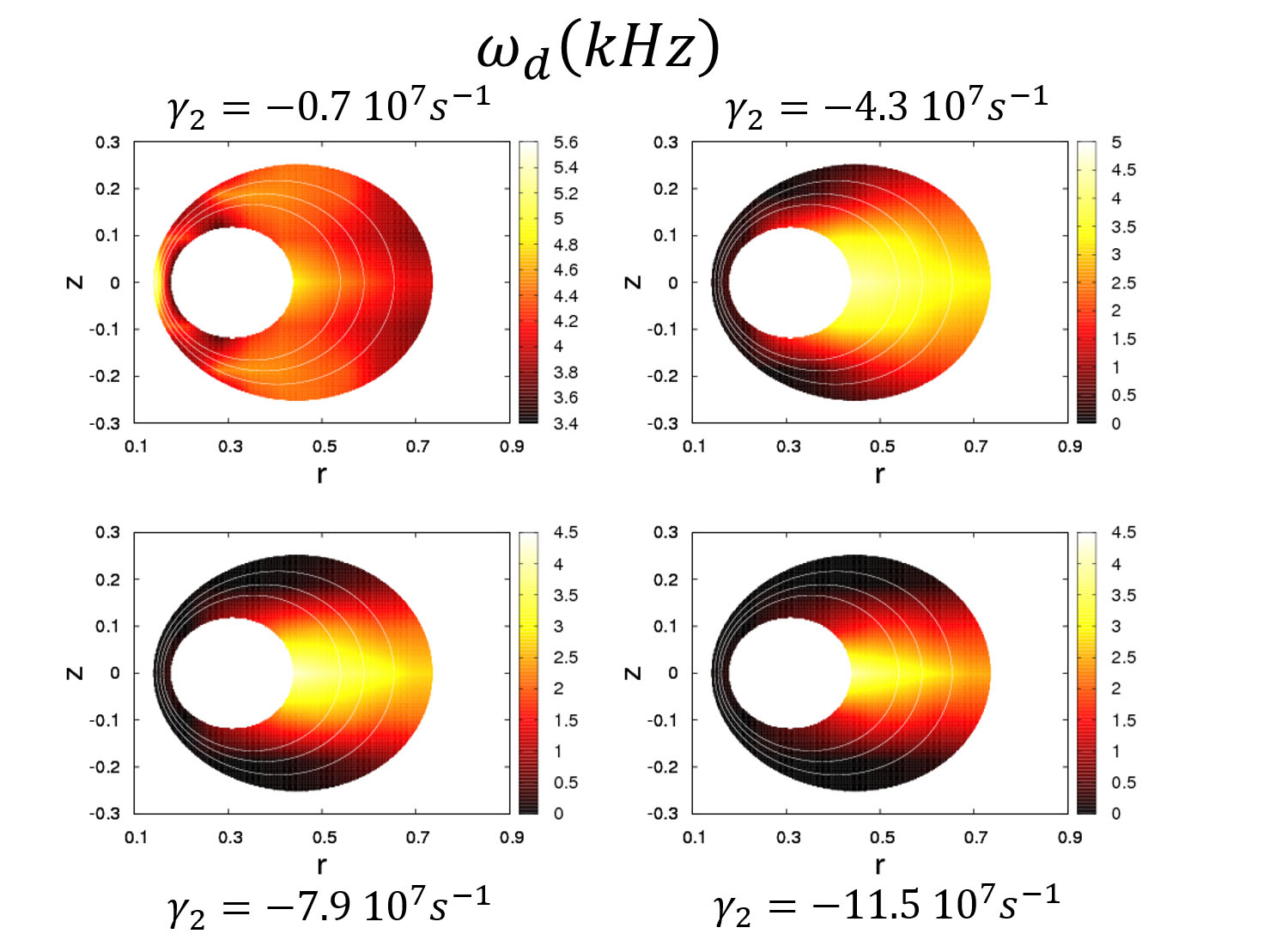}
 \caption{Plots of the angular frequency $\omega_{d}(kHz)$ of rotation for four different values of $\gamma_{2}(s^{-1})$.}
 \label{omegadg2}
 \end{figure}

\begin{figure}[h]
 \hspace*{-0.5cm}
 \includegraphics[scale=0.185]{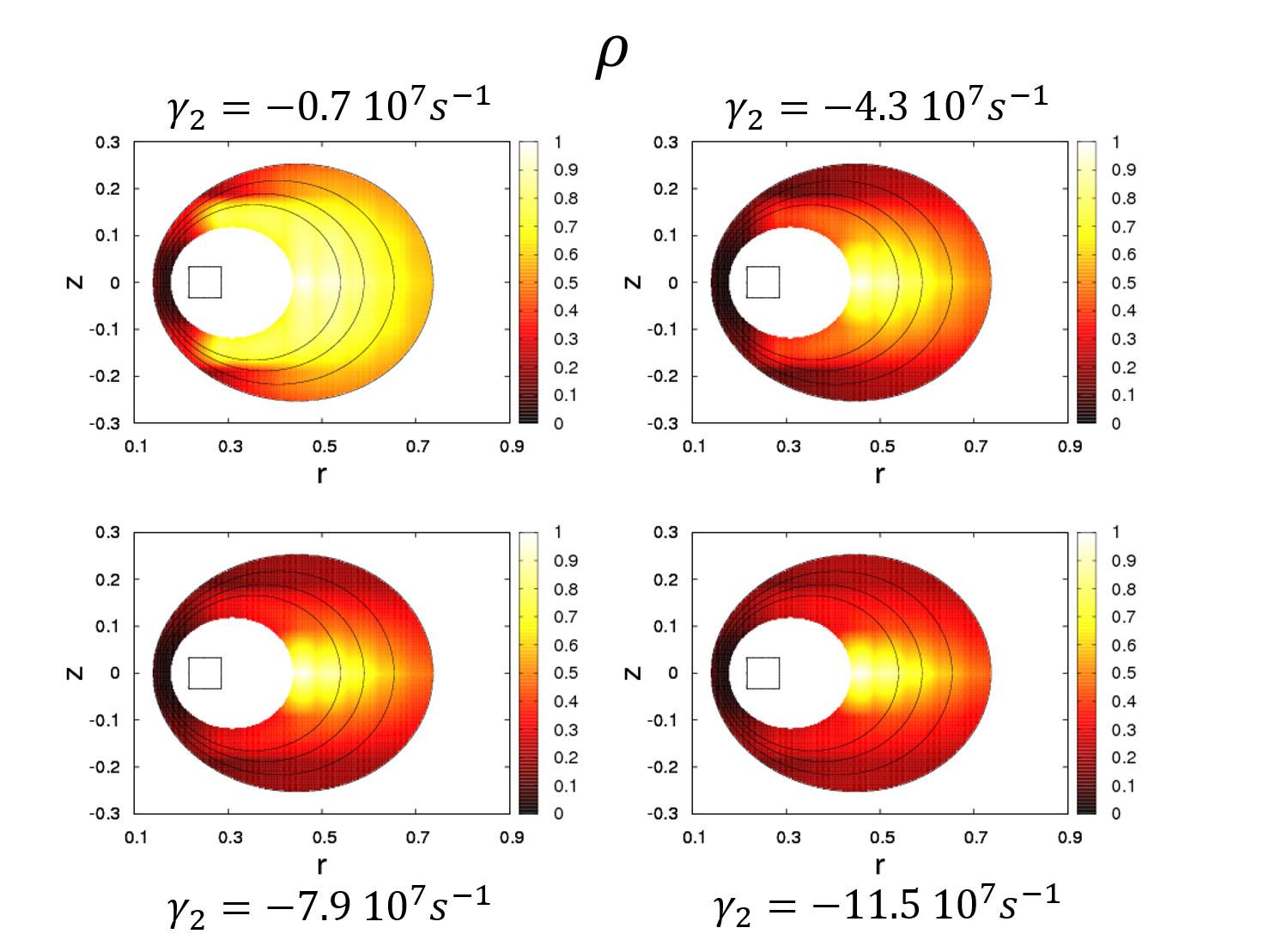}
 \caption{Plots of the density $\rho$ (normalized) for four different values of $\gamma_{2}(s^{-1})$.}
 \label{rhog2}
 \end{figure}

 The chemical potential $\gamma_{2}$ of the bounce action $J_\parallel$
 has a rather strong influence on the equilibrium distribution;
 see Figs. \ref{omegadg2}  (plots of $\omega_d$) and \ref{rhog2}  (plots of $\rho$).
 With the optimum value $\gamma_{2}=0.7\times10^{7}\,s^{-1}$, the density $\rho$
 has a broad distribution with a highly homogeneous drift frequency $\omega_d$.
 However, when $\gamma_{2}$ is increased, the density profile shrinks to a disk-like shape
 with a strong shear in $\omega_{d}$.
 Here, other parameters are fixed at  $\phi_{0}=-250\,V$, 
  $\beta^{-1}=50$eV, and $(\gamma_{1},\gamma_{3})=(-0.795\times10^{-1}\,T,1.25\times10^{4}\, Wb^{-1}\,eV)$.
 The change of $\chi$ as a function of $\gamma_{2}$ is given in Fig. \ref{dependences}.



\subsection{Chemical potential $\gamma_{3}$ of the angular momentum $\psi$}
\begin{figure}[h]
 \hspace*{-0.5cm}
 \includegraphics[scale=0.185]{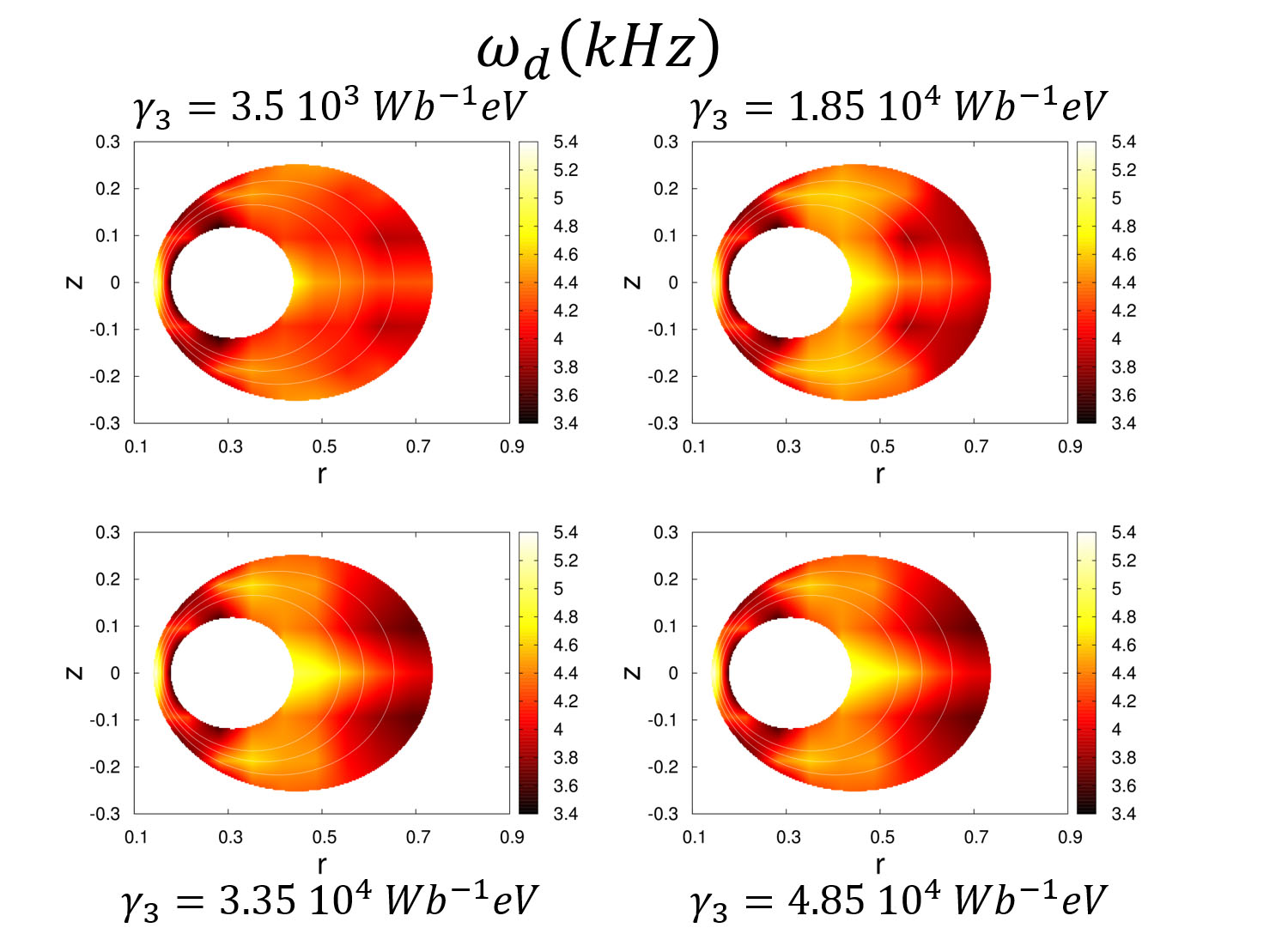}
 \caption{Plots of the angular frequency $\omega_{d}(kHz)$ of rotation for four different values of $\gamma_{3}(Wb^{-1}\,eV)$.}
 \label{omegadg3}
 \end{figure}

\begin{figure}[h]
 \hspace*{-0.5cm}
 \includegraphics[scale=0.185]{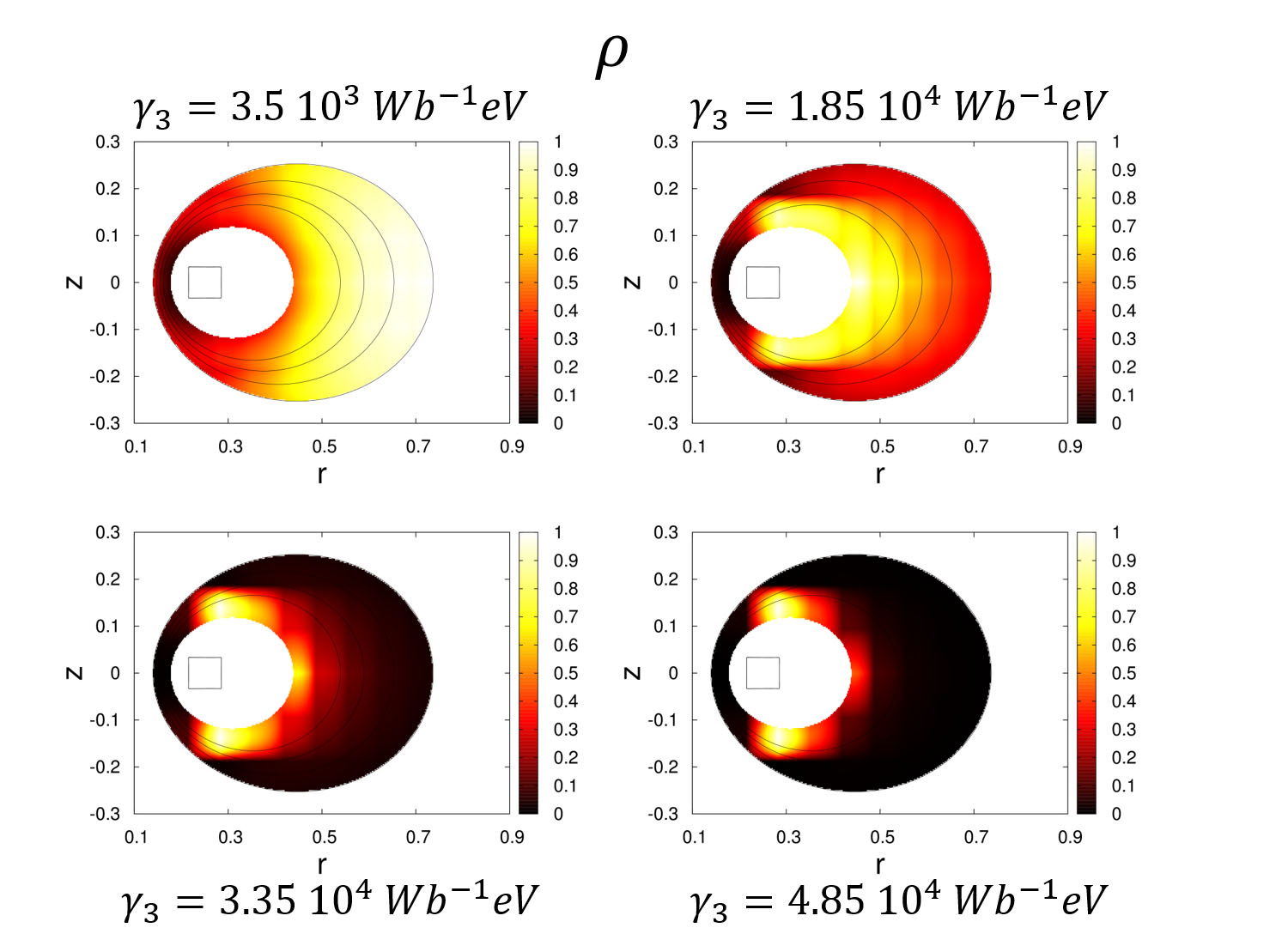}
 \caption{Plots of the density $\rho$ (normalized) for four different values of $\gamma_{3}(Wb^{-1}\,eV)$.}
 \label{rhog3}
 \end{figure}

 The chemical potential $\gamma_{3}$ of the angular momentum $\psi$
 also has a strong influence on the equilibrium;
 see Figs. \ref{omegadg3}  (plots of $\omega_d$) and \ref{rhog3}  (plots of $\rho$).
 As $\gamma_{3}$ is increased, the distribution function becomes $f \approx e^{\beta(\gamma_{3}\psi-q\phi)}$,
 and then the drift frequency scales as $\omega_{d}\sim r^{-1}$, since $\omega_{d}=q\partial\phi/\partial\psi$
 and, at $z=0$, $\psi\sim r^{-1}$ and $\phi\sim r^{-2}$ (if we assume a constant particle number per flux tube volume \cite{hasegawa2005}, $\rho\sim r^{-4}$ which gives $\Delta\phi\sim r^{-4}$).
 Here, other parameters are fixed at  $\phi_{0}=-250\,V$, 
 $\beta^{-1}=50$eV, and $(\gamma_{1},\gamma_{2})=(-0.795\times10^{-1}\,T,-0.72\times10^{7}\,s^{-1})$. 
The change of $\chi$ as a function of $\gamma_{3}$ can be found in Fig. \ref{dependences}.


\subsection{Inner boundary potential $\phi_{0}$}
\begin{figure}[h]
 \hspace*{-0.5cm}
 \includegraphics[scale=0.185]{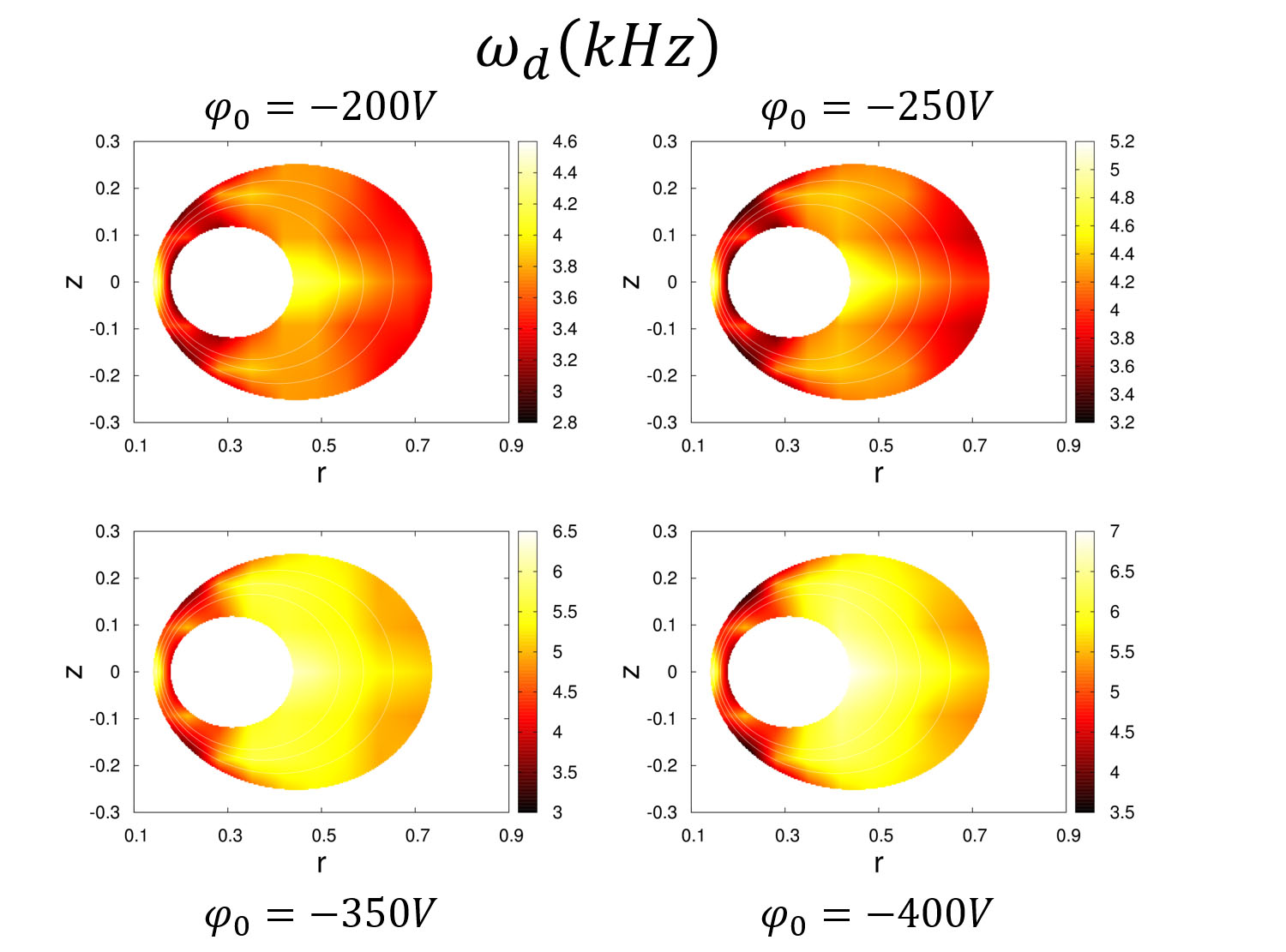}
 \caption{Plots of the angular frequency $\omega_{d}(kHz)$ of rotation for four different values of $\phi_{0}(V)$.   }
 \label{omegadphicoil}
 \end{figure}

\begin{figure}[h]
 \hspace*{-0.5cm}
 \includegraphics[scale=0.185]{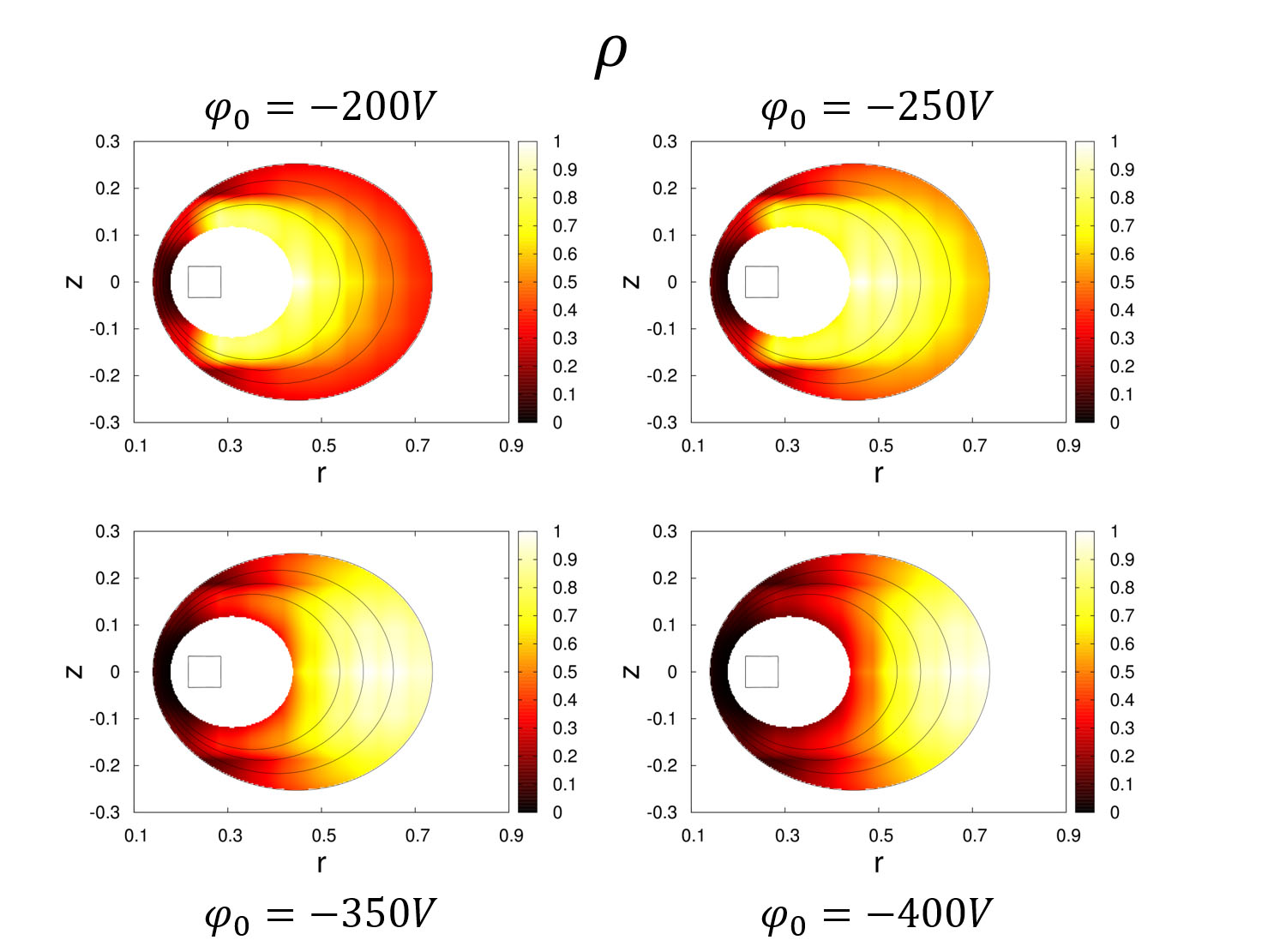}
 \caption{Plots of the density $\rho$ (normalized) for four different values of $\phi_{0}(V)$.   }
 \label{rhophicoil}
 \end{figure}

In addition to the four parameters $\beta$, $\gamma_1$, $\gamma_2$, and $\gamma_3$, there is another important parameter $\phi_{0}$
that is the boundary value of the potential $\phi$ at the surface of the internal magnet.
By changing $\phi_{0}$, the confinement is dramatically improved; see the experimental result reported in~\cite{Saitoh2004}.
On the levitated magnet experiment~\cite{yoshida2010}, however,
$\phi_{0}$ is spontaneously determined, because the coil surface is floating.
Figures \ref{omegadphicoil} and \ref{rhophicoil}, respectively, show the
distribution of the drift frequency $\omega_{d}$ and the density $\rho$
for four different values of $\phi_{0}$.
Other parameters are fixed at $\beta^{-1}=50$eV and
 $(\gamma_{1},\gamma_{2},\gamma_{3})=(-0.795\times 10^{-1}\,T,-0.8 \times 10^{7}s^{-1},1.35\times10^{4}\,Wb^{-1}\,eV)$. 
Interestingly, as $\phi_0$ changes, the confinement region (the density clump) moves, while
the rigidity of the rotation is not influenced by $\phi_0$; see Fig.\,\ref{dependences}.

\section{Conclusion}
 Putting a classical knowledge into a wider perspective, a deeper principle may emerge;
 here we have described such an example of paradigm shift in the study of the \emph{relaxed state} (or, \emph{thermal equilibrium}) of charged particles.
Formulating a relaxed state on a topologically constrained macroscopic phase space,
we have found that a rigidly-rotating equilibrium can self-organize even in an inhomogeneous magnetic field.
In an experimental system, tuning of the parameters $\gamma_1,\gamma_2,\gamma_3$, and $\phi_{0}$
occurs spontaneously through the relaxation process.
Among them, $\gamma_2$ and $\gamma_3$ influence strongly on the profile of $\omega_d$.
Whereas we formulated the equilibrium assuming that the corresponding $C_{J_\parallel}$ and $C_\psi$ are given
(then, the Lagrange multipliers $\gamma_2$ and $\gamma_3$ are determined by prescribed $C_{J_\parallel}$ and $C_\psi$),
the plasma may change them (by dissipating the constants of motion) in order to relax into the thermal equilibrium;
the self-organization is a process that selects optimum $C_{J_\parallel}$ and $C_\psi$ (hence, $\gamma_2$ and $\gamma_3$).
In fact, these two actions are relatively ``fragile'' with respect to the other constant $C_\mu$.

 The present model of relaxed states differs from the previously formulated
 Boltzmann distribution of a neutral plasma\,\cite{yoshida2013},
 which is created by constraining only $C_\mu$ and $C_{J_\parallel}$
 (in addition to the standard constraints $N$ and $E$) in maximizing the entropy;
 freeing $C_\psi$ means that $\psi$ is not a conserved quantity.
 In the present formulation, however, we also constrain $C_\psi$, while we demand $\partial f/\partial\psi=0$ as the criteria of the relaxed state.
 The latter is a weaker criterion,
 i.e., the present solution is not necessarily the maximum entropy state of the former setting, 
 as far as $\psi$ is freed from the kinetic energy (i.e., we omit the energy of the drift velocity)
 and is deemed as a spatial coordinate variable.
 This is obvious by putting $\gamma_3$ (the Lagrange multiplier on $C_\psi$) $=0$ in
 the thermal equilibrium (\ref{Penning-2}).
 Then, the solution becomes $\omega_d\rightarrow0$, $\rho\rightarrow0$ with $\beta q\phi\rightarrow$ constant ($=\infty$),
 i.e., the radius of the plasma column diverges, implying \emph{no confinement}..
 The constraint on the total angular momentum $C_\psi$ yields a finite radius of confinement.

\section*{Acknowledgments}
The authors acknowledge the stimulating discussions and suggestions of Professor A. Hasegawa, Professor S. M. Mahajan, and Professor R. D. Hazeltine.
This work was supported by the Grant-in-Aid for Scientific Research 
(23224014) from MEXT-Japan.


\end{document}